# Phyllotaxis: a Model

F. W. Cummings*
University of California Riverside (emeritus)

**Abstract**
 A model is proposed to account for the positioning of leaf outgrowths from a plant stem. The specified interaction of two signaling pathways provides tripartite patterning. The known phyllotactic patterns are given by intersections of two 'lines' which are at the borders of two determined regions. The Fibonacci spirals, the decussate, distichous and whorl patterns are reproduced by the same simple model of three parameters.

*present address: 136 Calumet Ave., San Anselmo, Ca. 94960; fredcmgs@berkeley.edu

# 1. Introduction

Phyllotaxis is the regular arrangement of leaves or flowers around a plant stem, or on a structure such as a pine cone or sunflower head. There have been many models of phyllotaxis advanced, too numerous to review here, but a recent review does an admirable job (Kuhlemeier, 2009). The lateral organs are positioned in distinct patterns around the cylindrical stem, and this alone is often referred to phyllotaxis. Also, often the patterns described by the intersections of two spirals describing the positions of florets such as (e.g.) on a sunflower head, are included as phyllotactic patterns. The focus at present will be on the patterns of lateral outgrowths on a stem. The patterns of intersecting spirals on more flattened geometries will be seen as transformations of the patterns on a stem. The most common patterns on a stem are the spiral, distichous, decussate and whorled. The spirals must include especially the common and well known Fibonacci patterns. Transition between patterns, for instance from decussate to spiral, is common as the plant grows.

# 2. The Pattern Model

The pattern model will be assumed to be based on the simplest version of the interaction of two signaling pathways. An activated signaling pathway is called an "Asp", and the smaller region separating two activated signaling pathways will be called the "Margin" region. In each growth cycle, the pattern will be then tripartite, consisting of two to-be determined Asp regions and a Margin region.

A tripartite pattern spontaneously emerges at each growth cycle due to the assumed action of the cell upon being activated by its specific ligand. This is simply described as two subsequent actions of each Asp:

1) Upon activation by its specific ligand, the Asp will lead to emission of further ligands of the same type (e.g., #1) into the extracellular space.

2) At the same time that the Asp is causing emission of like ligand, it also acts to prevent such emission of ligand of type two, the ligand of the second signaling pathway.

The same actions as in 1) and 2) apply to both signaling pathways involved in the interaction. For both actions 1) and 2), this occurs in a way proportional to the density of the two Asps in a given small region of the epidermal sheet. The two Asp densities are denoted as $R_1$ and $R_2$, while the corresponding ligand densities in the same small region will be denoted as $L_1$ and $L_2$. Then we can write

$$\partial L_1 / \partial t = D_1 \nabla^2 L_1 + \alpha R_1 - \beta R_2, \qquad 2.1)$$

with a similar equation for ligand (protein) number two, namely

$$\partial L_2 / \partial t = D_2 \nabla^2 L_2 - \alpha R_1 + \beta R_2, \qquad (2.2)$$

where the rates α and β are similar. The equations with α and β set to zero are the familiar diffusion equations. The "Laplace" operator, roughly speaking, denotes the difference between the in-flux of ligand from an average of its neighbors into a given small region, minus the out-flux to the same neighbors. The diffusion rates $D_1$, $D_2$ are generally about the same magnitude, perhaps even equal. Thus this is not a "Turing" reaction-diffusion model.

A reasonable (and simplest) assumption is that $R_1$ and $R_2$ are each proportional to their respective ligand densities in any small region. The receptors and their ligands have high affinity for each other, so that the number of active receptors is reasonably proportional to the ligand in the same small space region at a given time. Then

$$\gamma_1 R_1 = L_1, \quad \text{and} \quad \gamma_2 R_2 = L_2. \qquad (2.3)$$

Clearly more complicated equation can easily be written (Cummings, 2006, 2009), but our purpose here is to emphasize concepts. Equations (2.1) and (2.2) can then be written solely in terms of the activated receptor densities $R_1$ and $R_2$. Nonlinearities must ultimately enter. In this simple model, they are assumed to to provide an upper cutoff for the R's (and L's). A particular form will not be of interest here; the shape of the Asps may be affected but the positions of the Margin regions separating the two different Asp regions will be largely unaffected by the constraining nonlinearity. Our main interest here is to illustrate the concept of how patterns of gene activity may form due to two simple actions of cells. Such actions are as yet to be found experimentally, however.

The question of stability can be examined in the usual way, by introducing an exponential space and time behavior (Murray, 1990). Then we take both activated receptor densities as $R_{1,2}$ proportional to exp(st+κ·x), and inquire under what conditions or parameter values the eqs. (2.1), (2.2) and (2.3) give rise to exponentially growing solutions in time, i.e., where "s" is positive. This leads, after solution of a quadratic equation, to the condition

$$\kappa^2 < k_1^2 + k_2^2 \equiv k^2. \qquad (2.4)$$

Spontaneous activation of the pattern from zero density occurs when the space is available, and a pattern does not form when the tissue is below a minimal size. The assumption of a lower density threshold for each pattern determines the width of the Margin region, providing a line-like region, and defining a line (the bisector) dividing the Margin region into two parts.

Using the definitions $k_1^2 = \alpha/(\gamma_1 D_1)$, $k_2^2 = \beta/(\gamma_2 D_2)$ and $f = \beta/\alpha$ allows he steady state version of eqs. (2.1), (2.2) and (2.3) to be written in the form

$$\nabla^2 R_1 + k_1^2 (R_1 - f R_2) = 0, \qquad (2.5)$$

and

$$\nabla^2 R_2 - (k_2^2 / f)(R_1 - f R_2) = 0. \qquad (2.6)$$

The simplest example of a tripartite pattern is the axially symmetric pattern on a sphere. In this case the pattern may emerge only after a critical radius $R_0$ is reached by growth, and is given by $k^2 R_0^2 = 2$. Observation of $R_0$ then allows determination of $k^2$. The sphere solutions can be easily verified to be, in the case that $k_1 = k_2$,

$$R_1 = C \cdot \text{Cos}^2(\theta/2), \quad \text{and } f \cdot R_2 = C \cdot \text{Sin}^2(\theta/2). \qquad (2.7)$$

Here C is ~ a constant maximum amplitude, which can be normalized away, and C=1. The angle θ is the usual polar angle, zero at the 'north pole' and π at the south pole. Figure 1 shows the two separate Asp regions as red and blue, each with a maximum amplitude at opposite poles, a minimum at the Margin region, and with the Margin region shown as a white band; a Margin 'line' will circle the equator, with θ = π/2.

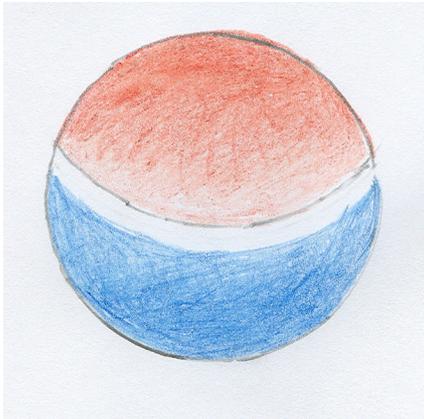

**Figure 1**
The tripartite spherically symmetric solution

The two regions, red and blue, each correspond to a particular genetic network activation. The Median region separating the two distinct regions correspond to a "toti-potent" region, which in the case of animals may be termed a "stem cell" region. If we may view the spherical epidermal cells of Figure 1 as a many-celled plant embryo (Lyndon, 1990), Figure 1 may model the earliest but multi-cellular separation of the plant cells into meristem (red) and root (blue) regions. Polarity is then provided, and arises spontaneously. Given the fact that the model always subdivides each original area again into binary parts upon further growth, one expects the spherical symmetry to be broken with the appearance of a Cos(φ) dependence (and multiplying a complex theta functional dependence, dependent on the geometrical shape, where the angle 'φ' circles the polar axis. Thus, intersections of Margin regions must occur, and these intersections will provide the positions ('points') of the first leaf primordia.

The next section examines the patterns of intersecting Margin regions that are allowed on a cylinder.

# 3. Phyllotaxis on a stem

The patterns occurring on the stems of plants (Mitchison, 1972; Douady and Couder, 1991; Green, 1996; Cummings and Strickland, 1998; Kuhlemeier, 2007) is presented in light of the present model. The suggestion is that, even though the biochemicals involved in the two kingdoms, animal and plant, are expected to be quite different, the simple tripartite patterning model (two Asps separated by Margin cells) of the previous section is assumed to have applicability to plants as well as animals ( Cummings (2009)). It is shown that while the present patterning model gives the known arrangements of plant patterns ("phyllotaxis"), it nevertheless explicitly forbids the number 'four' in a spiral arrangement with one 'leaf' on each level. This means that, counting up or down the stem from some initial leaf, (taken as at the origin) one leaf per level, the total number of leaves until a repeat (a leaf directly above the initial leaf) occurs, **'four '** is forbidden by the present pattern algorithm. In fact, what is observed in nature is that the number four is strikingly less frequent in this case than the numbers 2, 3, 5 or 8, which are Fibonacci numbers. This argues the case for bias in development, albeit now for plant rather than animal development. One is challenged to find a plausible argument from natural selection or adaptation for such paucity of 'four' as is observed to occur. There is clearly no structural reason for the absence of spiral 'four'.

A review of phyllotaxis has been given recently (Kuhlemeier, 2007), providing much insight into the molecular basis of this subject. Auxin plays a particularly key role (Reinhardt et al., 2003). A complex quantitative analysis, taking into account the known molecular components, has been given by Smith et al (Smith et al., 2006). The present model hopes to augment these contributions although from a quite different (and simpler) point of view.

Outgrowth of leaves from a common stem will occur from points of intersection of two Margin regions. A prediction (perhaps better: a 'suggestion') is that a "master regulatory" gene, will be found at these stem positions; in the animal case, outgrowths from the main body are accompanied by activation of the distal-less gene. Leaf locations are shown in Figure 2 as Margin intersections for the spiral case of N= 3, 5 and 8. In Figure 2, the cylinder has coordinates $x/x_0$, $y/y_0$ with circumference $x_0$, and stem repeat length $y_0$.

The simple patterning mechanism of the present paper in fact reproduces all Fibonacci spiral patterns (Mitchison, 1972; Douady and Couder, 1991; Cummings and Strickland, 1998), as well as the common decussate and distichous patterns, and numerous others commonly seen, such as whorls. Decussate patterns are very common, and consist of two leaves at the same level, placed 180° apart on the stem, with the next pair up (or down) along the stem rotated related to this first pair by 90°. Decussate patterns are the simplest example of alternating 'whorl' patterns, which may have three, four, etc., leaves at the same level. Distichous phyllotaxis is also very common, where successive higher (or lower) single leaves are 180º from the preceding one, examples being corn, ginger and ferns. Superposed whorls also commonly occur, where, in the most common case, a pattern of two leaves situated 180º at the same level, followed at another level by two leaves directly above the original pair. This latter pattern is particularly common in compound leaves. All of these patterns are easily reproduced by the model of Section 2.

Solutions to the model of Section 2 as applied to plant patterns on a stem are given. A new pair of integers is introduced by the model, the pair (p, q) designating a given pattern. The (p, q) pair underlies and predicts the more usual 'parastichy" integer pair (m, n) (e.g., Mitchison, 1972).

The unit square is bounded by the coordinates $x/x_o$ and $y/y_o$ in Figure 2. A positive integer pair (p, q), $p \geq q$, designates a particular pattern. The axial coordinate around the stem is $x/x_o$, with $x_0$ being the stem circumference, so that all points at $x = 0$ and $x = x_o$ are the same. The normalized length up (or down) the stem is taken to be $y/y_o$, and the horizontal line $y = 0$ has the same values as $y = y_o$. Interest is focused on a repeating pattern in the coordinates, with the boundary conditions that y=0 occurs at the point $x/x_o = 0$ or 1. The solutions of eqs. (2.4), (2.5) and (2.6) have been constructed so that there is always an intersection of Margin ($R_1 = R_2$) at the origin $x/x_o = 0$ and $y/y_o = 0$, (i.e., the point (0,0)), as well as at the other three corners (1,0), (1,1), and (0,1). The repeating leaf at (0, 1) is not counted in the total number 'N' of leaves in a pattern.

The solutions to eqs. (2.5) and (2.6) can be written in terms of the variables $\theta_1(x,y)$ and $\theta_2(x,y)$ as

$$R_1 = (2 + \sin(\theta_1) + \sin(\theta_2)), \qquad (3.1a)$$

and

$$fR_2 = (2 - \sin(\theta_1) - \sin(\theta_2)). \qquad (3.1b)$$

Here $\theta_1$ and $\theta_2$ are given by

$$\theta_1 = 2\pi(px/x_0 + qy/y_0), \quad \text{and} \quad \theta_2 = 2\pi(qx/x_0 \pm py/y_0). \qquad (3.2)$$

with

$$k^2 A = (2\pi)^2(p^2+q^2)(y_0/x_0+x_0/y_0), \qquad (3.3)$$

The area of a single repeat along the stem is $A = x_0 y_0$. Equation (3.3) shows that if the stem radius is kept constant as growth occurs and A increases, then there need be no change in pattern (p, q). On the other hand, if $x_0$ and $y_0$ increase proportionally with growth, then the phyllotactic pattern (p, q) must change. Both are observed.

Intersecting Margin lines (defined as $R_1 = R_2$) denote the positions of leaves or florets, and are located by requiring that the argument in each Sine function of eqs. (3.1a, b) be $2\pi$ times an integer 'i' or 'j' in each case. This gives at once the two equations, from eqs. (3.1a, b),

$$y/y_o = (p/q)x/x_o + i/q, \qquad (3.4)$$

$$y/y_o = \pm(q/p)x/x_o + j/p. \qquad (3.5)$$

A **given leaf** is designated by a particular integer pair (i, j) designating an intersection of the two straight lines within the rectangle of Figure 2. **A particular pattern is specified by an integer pair 'p, q'**. Equations (3.4) and (3.5) are in a standard form. The two straight lines have **slopes $S_1 = (p/q)$ and $S_2 = \pm(q/p)$,** and the terms $i/q$ and $j/p$ are intercepts of the straight lines with the $y/y_0$ axis. The y intercepts do not all lie within the unit square. Pattern construction follows at once from eqs. (3.4) and (3.5). The intersections of the lines shown in Figure 2 gives the positions of the leaves in a given pattern specified by the pair (p, q). The number of leaves in a given pattern, in the case that there is only one leaf per level of either x or y, and (p, q) are relatively primed, is given by the expression

$$N(p, q) = p^2 \pm q^2. \qquad (3.6)$$

This can be seen by eliminating x (or y) in eqs. (3.4) and (3.5) and observing that $0 < y/y_o \leq 1$, which implies that the maximum number of leaves in a pattern is given by eqs. (3.4) and (3.5). In the case of the plus sign in eqs. (3.4) and (3.5) the two sets of straight lines, corresponding to a set of integers 'i' in the one case and 'j' in the other, have opposite slopes, while for the negative sign in 'N' the two sets of straight lines both have the same slope. This is shown in Figure 2 for the cases of $N = 3$, $N = 5$ and $N = 8$.

When there are 'J' leaves on the same level, the expression for the number of leaves in a pattern becomes simply $N(p, q, J) = (p^2 \pm q^2)/J$. A 'whorl' pattern has $p = q$, the most common example being the decussate pattern, when $p = q = 2$, $J=2$ and $N = 4$.

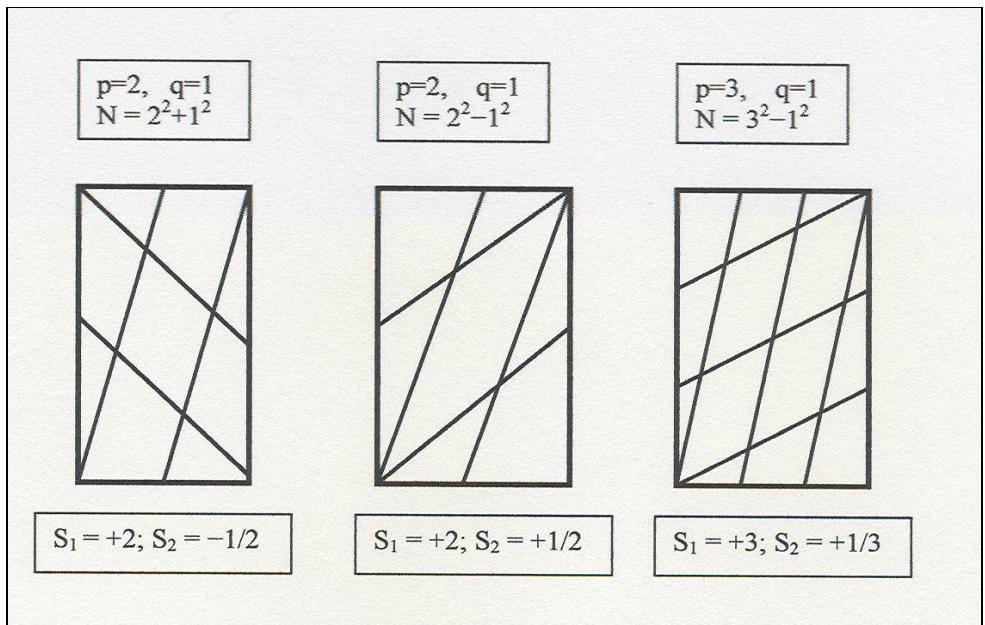

**Figure 2**
The Margin lines are shown, where $R_1 = R_2$ of eqs. (3.1a) and (3.1b) are equal. The slopes are given by $S_1 = p/q$ and by $S_2 = \pm q/p$. The horizontal coordinates are $x/_0$ and vertical $y/y_0$. Left to right, $N = 5$, $N = 3$, and $N = 8$.

In the case of the spiral with one leaf per level, growth considered as **transition** from one Fibonacci pattern to the next **occurs simply by adding one** **leaf to each existing row.** The plus/minus sign then alternates in N of eq. (3.6), and p or q is alternately increased to the next value, as shown in Table 1. There are $R_+ = p^2 - (p-q)^2$ pattern rows in the case of the plus sign, and $R_- = q^2 + (p-q)^2$ rows in the case of the minus sign in the expression for N in eq.(3.6). The m's correspond to the usual designation of parastichies, where N = m + n. Figure 5 illustrates three Fibonacci pattern spirals, $N = 3 = 2^2 - 1^2 = 2 + 1$, and $N = 5 = 2^2 + 1^2 = 3 + 2$, and $N = 8 = 3^2 - 1^2 = 5 + 3$.

Whorl patterns have p = q, so that N = 2p, and J = p, in the formula $N = (p^2 \pm q^2)/J$. Superposed whorl patterns have p > q = 0, when one set of lines has slope zero, and the other set has infinite slope. The common superposed whorl with p = 2, q = 0, and N = p, has two leaves on the same level (J = p = 2) displaced by 180°, with a superposed pair directly above. Compound leaves usually display such a pattern.

There is a simple analytical transformation taking the unit square into an annulus, while preserving the form of the model equations. The coordinate 'y' is mapped into the polar coordinate 'r' while 'x' maps into the polar angle 'θ'. The lines y = 0 and $y = y_o$ are then mapped into two concentric circles, while the lines x = 0 and $x = x_o$ are mapped into the straight lines representing the angles 0 and 2π in the plane. In the case that x maps into an angle less than 2π, the square maps into a conical figure rather than an annulus. The two sets of intersecting straight lines of eqs. (3.1) and (3.2) are in either case mapped into two sets of intersecting logarithmic spirals.

What is clear from eq. (3.6) is that the number '**four**' is not included among spiral patterns, those with a single leaf on a level, according to the model. Such is also very rare in nature. No adaptive or selective reason for this is forthcoming. Rather, it is a result of the particular pattern formation algorithm of the present paper.

The Table 1 shows increasing Fibonacci pattern numbers determined as an alternating increase in the two basic determining parameters 'p' and 'q'. The more usual parastichy numbers 'n' and 'm' are shown on the right of the Table 1.

| N | $p^2 \pm q^2$ | m+n |
|---|---|---|
| 2 | $1^2+1^2$ | 1+1 |
| 3 | $2^2-1^2$ | 2+1 |
| 5 | $2^2+1^2$ | 3+2 |
| 8 | $3^2-1^2$ | 5+3 |
| 13 | $3^2+2^2$ | 8+5 |
| 21 | $5^2-2^2$ | 13+8 |
| 34 | $5^2+3^2$ | 21+13 |
| 55 | $8^2-3^2$ | 34+21 |
| 89 | $8^2+5^2$ | 55+34 |
| · | | |
| · | | |
| · | | |

Table 1

**Table 1:   The Fibonacci Pattern**

The table shows that as growth occurs, progression from one Fibonacci pattern to the next comes about by addition of one leaf to each row, as space allows. There are $R_- = q^2+(p-q)^2$ rows in the case of the negative sign in $N = p^2 \pm q^2$, when the two sets of straight lines have the same slopes, and $R_+ = p^2 - (p-q)^2$ rows in the case that the two sets have opposite slopes, and illustrated in Figure 2. The relatively primed integers p and q increase alternately with increasing area. The pair (p, q) underlie the more usual and larger 'parastichy' pattern designations (m, n) shown in the right-most column.

It may be briefly noted that it is to be expected that, starting from a radially symmetric embryo such as imagined in Figure 1, opposite leaf primordia will emerge when a critical area is reached, and the pattern will subsequently settle into a (e.g.) spiral pattern with stem growth. For example, consider a hemisphere or a cone; given that the present pattern algorithm of Section 2, the pattern must always give rise to two 'determined' Asp regions separated by a Margin region. Then the two sides of the cone or hemisphere will be designated as two different Asp regions upon further growth, outlined by two Margin regions: one Margin region will be a circle at the cone or hemisphere base, while the second will divide the surface into two (equal) parts, and (e.g.) go through the north pole. The two Margin line intersections then occur at the base, or periphery, and will determine the positions of two initial primordia.

## 5. Summary

The present argues for pattern formation agents different from the usual. The usual concept of 'morphogen' requires a long-range diffusing substance (Kerzsberg and Wolpert, 2007). The term '**Asp**' is used here to refer to the density of a particular activated signalling pathway, and thus to the density of its associated factor that activates its specific transcription factor. The term 'Asp' denotes a spatial region in which specific transcription factors have been activated in the nucleus, and where particular selector genes are activated (Gerhart and Kirschner, 1997). As such, no discussion of whether ligand diffuses around cells or through them is required, nor is there discussion of the method of achieving long range diffusion. Binary patterns are given at each growth cycle, providing successive overlapping patterns, and a possible unique genetic specification after each growth cycle. Pattern complexity increases with each cycle, a cycle specified by the growth and decay of a given signaling pair.

The Asps may be assumed to interact with morphogenetic movement; one Asp increases the apical/basal ratio, while the other decreases the difference (e.g., Cummings, 2006). Specific gene activation not specified here will give rise to specific cell shapes. Such epidermal sheet movements will necessarily act back to affect gene activation, but such feedback from gene to cell shape change has not been considered here. Nor have many other complications been included, in particular the effect on pattern of the plant cell wall.

The model hopes to introduce new possibilities for patterning of stem cells and designation of 'points' of plant stem outgrowth. As a realistic model it is clearly far too simple. There are many omissions that could be mentioned. The assumption that the inner tissues can be neglected in patterning is certainly one such omission; the relevant assumption has been made that all patterning takes place in the outer plant cell layer of the stem, and interaction with the inner tissues can be neglected. Relevant molecules such as PIN1 are located in the internal cells; the present model does not integrate the events of cells at the surface of the stem with those of the inner tissues. Initiation of pattern in the meristem has been mostly neglected here (Smith et al. 2006), but rather attention has been focused on the pattern of lateral outgrowth from the stem. No attempt has been made here to make contact with the bio-molecules of the suggested coupled signaling pathways that are the proposed source of the pattern. Many perhaps most of these are probably as yet unknown. It may be that the hormone auxin acts as a ligand, and the PIN proteins catalyzes the required export of the auxin from the cell (Petrasek et al., 2006). Auxin is required for proper positioning of organs, and auxin can induce lateral organ outgrowth (Reinhardt et al., 2000) by local application. In the present model two key ligands are required, and such have not been described as yet.

## Acknowledgement

I wish to express appreciation for encouragement and inspiration provided through many years by my friend and outstanding scientist Brian Goodwin **(1931-2009)**. I would also like to thank Barbara Picard PhD for her many helpful comments and corrections.